\documentstyle[aps,multicol,pra,tighten,eqsecnum,graphicx]{revtex}
\begin{document}

\draft

\title{Robust unravelings for resonance fluorescence}
\author{H.M. Wiseman and Zoe Brady}
\address{School of Science, Griffith University, Nathan, Brisbane, 
Queensland 4111 Australia. }
\maketitle

\begin{abstract}
Monitoring the fluorescent radiation of an atom unravels the master 
equation evolution by collapsing the atomic state into a pure state 
which evolves stochastically. A robust unraveling is one that 
gives pure states that, on average,
are relatively unaffected by the master equation evolution (which 
applies once the monitoring ceases). The ensemble of pure states 
arising from the maximally robust unraveling has been suggested to be 
the most natural way of representing the system [H.M. Wiseman and 
J.A. Vaccaro, Phys. Lett. A {\bf 250}, 241 (1998)]. We find that the 
maximally robust unraveling of a resonantly driven atom 
requires an adaptive interferometric 
measurement proposed by Wiseman and Toombes 
[Phys. Rev. A {\bf 60}, 2474 (1999)]. The resultant ensemble consists 
of just two pure states which, in the high driving limit, are 
close to the eigenstates of the driving Hamiltonian 
$\Omega\sigma_{x}/2$. This ensemble is the closest thing to a 
classical limit for 
a strongly driven atom. We also find that it is possible to reasonably 
approximate this ensemble using just homodyne detection, an example 
of a continuous Markovian unraveling. This has implications for other 
systems, for which it may be necessary in practice 
to consider only continuous 
Markovian unravelings.
\end{abstract}  

\newcommand{\beq}{\begin{equation}} 
\newcommand{\eeq}{\end{equation}}
\newcommand{\bqa}{\begin{eqnarray}} 
\newcommand{\eqa}{\end{eqnarray}}
\newcommand{\nn}{\nonumber} 
\newcommand{\nl}[1]{\nn \\ && {#1}\,}
\newcommand{\erf}[1]{Eq.~(\ref{#1})}
\newcommand{\erfs}[2]{Eqs.~(\ref{#1})--(\ref{#2})}
\newcommand{\dg}{^\dagger}
\newcommand{\rt}[1]{\sqrt{#1}\,}
\newcommand{\smallfrac}[2]{\mbox{$\frac{#1}{#2}$}}
\newcommand{\half}{\smallfrac{1}{2}} 
\newcommand{\bra}[1]{\langle{#1}|} 
\newcommand{\ket}[1]{|{#1}\rangle}
\newcommand{\ip}[2]{\langle{#1}|{#2}\rangle}
\newcommand{\norm}[1]{\ip{#1}{#1}}
\newcommand{\op}[2]{\ket{#1}\bra{#2}}
\newcommand{\proj}[1]{\op{#1}{#1}}
\newcommand{\sch}{Schr\"odinger } 
\newcommand{\schs}{Schr\"odinger's }
\newcommand{\hei}{Heisenberg } 
\newcommand{\heis}{Heisenberg's }
\newcommand{\bl}{{\bigl(}}
\newcommand{\br}{{\bigr)}} 
\newcommand{\ito}{It\^o }
\newcommand{\str}{Stratonovich } 
\newcommand{\dbd}[1]{\frac{\partial}{\partial {#1}}}
\newcommand{\sq}[1]{\left[ {#1} \right]}
\newcommand{\cu}[1]{\left\{ {#1} \right\}}
\newcommand{\ro}[1]{\left( {#1} \right)}
\newcommand{\an}[1]{\left\langle{#1}\right\rangle}
\newcommand{\implies}{\Longrightarrow}
\newcommand{\ve}{\varepsilon}

\begin{multicols}{2}

\section{Introduction}

Some states of open quantum systems are more robust than others. That is, 
they are less perturbed by the system dynamics. This fact  has been the 
subject of a long-running and active research program 
\cite{Gea90,Zur93,ZurHabPaz93,Gal95,BarBurVac96,ParScu98,Gea98}. 
Recently, one of us 
and Vaccaro have introduced into this program a formalism with a 
number of distinctive features \cite{WisVac98,WisVac00}. 
This formalism consists of  finding the 
maximally robust unraveling (MRU) for the open quantum system. Its 
introduction was  motivated by a desire to better understand the rich 
dynamics of open quantum systems in general \cite{WisVac98}, 
and that of the atom laser in 
particular \cite{WisVac00}.

The use of the term ``robust unraveling" rather than ``robust state" 
encapsulates two of the distinctive features of the work of 
Refs.~\cite{WisVac98,WisVac00}. An 
unraveling is a way of measuring the environment of an open quantum system 
such that the system state can be described by a pure state undergoing 
stochastic evolution.  This is always possible in principle if the 
unmonitored system obeys a Markovian master equation, which we will assume 
to have a unique stationary state. In the long time limit, the ``unraveled" 
system  will be in a pure state drawn at random from a particular ensemble 
of pure states defined by the unraveling. It is from the consideration of 
such an ensemble of pure states that the two distinctive features of our 
approach are met. The first is that it is not individual pure states  
whose robustness are to be calculated, but rather a whole {\em ensemble of 
pure states}, the  average robustness of which is calculated. The second 
is that the pure states in this ensemble are {\em physically realizable} 
in the sense that they are the states of the system known to an 
experimenter using the appropriate measurement scheme on the system's 
environment.

 As noted above, the principle application of the maximally robust 
unraveling formalism has been to a model for an atom laser (a continuously 
damped and replenished gaseous Bose-Einstein condensate). This work is a 
specialized application in two ways. First, the system itself has a high 
excitation number and so is a quantum system in the classical limit. 
Second, only a subset of the set of all possible unravelings was 
considered. This subset (which is still infinite) contains those 
unravelings that lead to continuous and Markovian  evolution of the 
system state vector \cite{note1}.  This restriction was necessary to make 
the problem tractable and was justified by the classicality of the system.

In this work we apply the formalism of MRU to a system with no classical 
limit (in the usual sense at least), a resonantly-driven fluorescent 
two-level atom.  
This system is one of the canonical examples of an open quantum system, 
and has surprisingly complex dynamics for its size. It is therefore worth 
investigating in its own right. But, even more importantly, it is simple 
enough that the maximally robust unraveling can be found analytically. It 
turns out that this MRU is neither continuous nor Markovian. This enables 
us to investigate the question of how closely one can approximate the 
ensemble of this MRU if one is restricted to considering continuous 
Markovian unravelings. The answer to this question has implications for 
the general usefulness of the MRU formalism, since  for typical systems it 
would be necessary to impose this restriction in order to make the 
formalism practical.

The structure of this paper is as follows. In Sec.~II we briefly review 
the MRU formalism. In Sec.~III we introduce the two-level atom model and 
derive an expression for the ensemble average survival probability (which 
is used to quantify the robustness) in terms of moments of the ensemble of 
state vectors. In Sec.~IV we look at one simple (but not maximally robust) 
unraveling, that resulting from direct detection of the atom's 
fluorescence, for comparison with other, more robust, unravelings. In 
Sec.~V we present the most robust unraveling and its ensemble of (in this 
case, just two) state vectors. In Sec.~VI we find the most robust 
unraveling from within the set of continuous 
Markovian unravelings. We compare this ensemble to the MRU of Sec.~V 
in Sec.~VII. We conclude with a discussion of the implications of 
our results in Sec.~VIII.

\section{Maximally Robust Unravelings}

\subsection{The Master Equation}

Open quantum systems generally become entangled	with their	
environment, and this causes their state to	become mixed.	
In many	cases, the system will reach an	equilibrium	mixed state	in the
long time limit. This is the sort of system	for	which our	
 approach to robustness,	
of finding the maximally robust	unraveling (MRU), can be applied	
without	modification.
	
If the system is weakly	coupled	to the environmental reservoir,
and	many modes of the reservoir	are	roughly	equally	affected by
the	system,	then one can make the Born and Markov approximations	
in describing the effect of	the	environment	on the system	
\cite{Gar91}. Tracing over (that is, ignoring) the state of	
the	environment	leads to a Markovian evolution equation	for	
the	state matrix $\rho$	of the system, known as	a {\em quantum
master equation}. The most general form	of the quantum master
equation that is mathematically valid is the Lindblad form	
\cite{Lin76}
\beq
\dot{\rho}=	-i[H,\rho] + \sum_{\mu=1}^M {\cal	D}[c_\mu]{\rho} \equiv	
{\cal L}\rho ,\label{genme}	
\eeq
where for arbitrary	operators $A$ and $B$,	
\beq \label{defcalD}
{\cal D}[A]B \equiv	ABA\dg - (A\dg AB + BA\dg A)/2 .
\eeq

If the master equation has a unique	stationary state (as we	will	
 assume	it does), then that	is defined by	
 \beq
{\cal L}\rho_{\rm ss} =	0	.
\eeq
This assumption	requires that ${\cal L}$ be	time-independent. In	
many quantum optical situations, such as resonance fluorescence,
one is	only interested	in the
dynamics in	the	interaction	picture, in	which the free evolution	
at optical frequencies is removed from the state matrix. Indeed,	
if one treats the driving field as classical, as we will do, 
it is	
necessary to move into such	an interaction picture in order	to
obtain a time-independent Liouvillian superoperator	${\cal L}$.
	
The	stationary state matrix	$\rho_{\rm ss}$	can	be expressed as	an	
ensemble of	pure states	as follows:	
\beq
\rho_{\rm ss} =	\sum_{k} \wp_{k} \ket{\psi_{k}}\bra{\psi_{k}},
\eeq
where the $\ket{\psi_{k}}$ are normalized state vectors 
and	the	$\wp_{k}$	are	
positive weights summing to	unity. The
(possibly infinite)	set	of ordered pairs,	
\beq
E =	\{ (\ket{\psi_{k}},\wp_{k}):k \},	
\eeq
we will	call an	ensemble $E$ of	pure states.
Note that there	is no restriction that the states be	
mutually orthogonal. This means	that
there are continuously infinitely many ensembles $E$ that represent
$\rho_{\rm ss}$.	
The	aim	of finding the MRU is to find the ``best''	or ``most natural''
representation for $\rho_{\rm ss}$.	
	
\subsection{Unravelings}	
	
As explained in	the Introduction above,	the	first criterion	for
our	most natural ensemble is that it be	physically realizable by	
monitoring the environment of the system. In the situation where	
a Markovian	master equation	can	be derived,	it is possible (in
principle) to continually measure the state	of the environment on	
a time scale large compared	to the reservoir correlation time but	
small compared to the response time	of the system. This
effectively	continuous measurement is what we mean by
``monitoring''.	In such	systems, monitoring	the	environment	does	
not	disrupt	the	system--reservoir coupling and the system will
continue to	evolve according to	the	master equation	if one	
ignores	the	results	of the monitoring.
	
By contrast, if	one	does take note of the results of monitoring
the	environment, then the system will no	
longer obey	the	master equation. Because the system--reservoir
coupling causes	the	reservoir to become	entangled with the	
system,	measuring the former's state produces information about
the	latter's state.	This will tend to undo the increase	in the
mixedness of the system's state	caused by the coupling.
	
If one is able to make perfect rank-one	projective (i.e. von
Neumann) measurements of the reservoir state, the system state
will usually be	collapsed towards a	pure state.	However	this is
not	a process that	itself can be described	by projective	
measurements on	the	system,	because	the	system is not being	
directly measured. Rather, the monitoring of the environment
leads to a gradual (on average)	decrease in	the	system's entropy.	
	
If the system is initially in a	pure state then, under perfect
monitoring of its environment, it will remain in a pure	state. Then	the
effect of the monitoring is	to cause the system	to change its pure	
state in a stochastic and (in general) nonlinear way. Such evolution
has	been called	a quantum trajectory \cite{Car93b},	and	can	be
described by a nonlinear stochastic	\sch equation for the system state 
vector	
\cite{DalCasMol92,GarParZol92,WisMil93c}. The nonlinearity	
and	stochasticity are present because they are a fundamental part of
measurement	in quantum mechanics. 

On average, the system still obeys the master equation. That is, if 
the increment in $\ket{\psi}$ under the SSE is $\ket{d\psi}$ then
\beq
{\rm E}[\ket{d\psi}\bra{\psi} + \ket{\psi}\bra{d\psi}+
\ket{d\psi}\bra{d\psi}] = {\cal L}{\rm E}[\ket{\psi}\bra{\psi}].
\label{ensav}
\eeq
Here ${\rm E}$ denotes the ensemble average	
with respect to	the	stochasticity of the SSE. This
stochasticity is evidenced by the necessity	of retaining the \ito term	
$\ket{d\psi}\bra{d\psi}$ \cite{Gar85}.
	
Because	the	ensemble average of	the	system still obeys the master	
equation, the 
stochastic \sch	equation) is said to {\em unravel} the	
master equation	\cite{Car93b}. It is now well-known	that there
are	many (in fact continuously many) different unravelings for a	
given master equation \cite{QSOSQO96}, corresponding to	different	
ways of	monitoring the environment.	
	
Each unraveling	${\cal U}$ gives rise to an ensemble of pure	
states
\beq \label{ens}	
E^{\cal	U} = \{ (\ket{\psi_k^{\cal U}}, \wp_k^{\cal	U}) :k\},	
\eeq
where $\ket{\psi_k^{\cal U}}$ are the possible pure states of the system	at	
steady state,
and	$\wp_k^{\cal U}$ are their weights. For master equations with	a unique	
stationary state $\rho_{\rm	ss}$,	
the	SSE is	ergodic	over $E^{\cal U}$ \cite{CrePC}
and	 $\wp_k^{\cal U}$	is
equal to the proportion	of time	the	system spends 
in state $\ket{\psi_k^{\cal U}}$.
The	ensemble $E^{\cal U}$ represents $\rho_{\rm	ss}$ in	that
\beq \label{decomp}	
\sum_{k} \wp_k^{\cal U} \ket{\psi_k^{\cal U}}\bra{\psi_k^{\cal U}} = \rho_{\rm ss},	
\eeq
as guaranteed by Eq.~(\ref{ensav}).

\subsection{Survival Probability}
	
Imagine	that the system	has	been evolving	
under a	particular unraveling ${\cal U}$	
from an	initial	state at time $-\infty$
to the stationary ensemble at the present time $0$.	
 It	will then be in	the	state $\ket{\psi_k^{\cal U}}$ with probability
$\wp_k^{\cal U}$.	If we now cease	to monitor the system
then  the state
will no	longer remain pure,	but	rather will	relax toward	
$\rho_{\rm ss}$
under the evolution	of Eq.~(\ref{genme}).	
	
This relaxation	to equilibrium will	occur	
at different rates for different states.	
For	example, some unravelings will tend	to	
collapse the system	into a pure	state that is very fragile,	in that	
it changes into a very different (and mixed) state
 as it relaxes to equilibrium. In this case	
the	ensemble would rapidly become a	poor representation
of the observer's expected knowledge
 about the system.	
Hence we can say that such an ensemble is a	``bad''	or
``unnatural'' representation of	$\rho$.	Conversely,	an unraveling	
that produces robust states	would remain an	accurate description	
for	a relatively long time.	We expect such a ``good'' or	
``natural''	ensemble to	give more intuition	about the dynamics of	
the	system.	The	most robust	ensemble we	interpret as the ``best''	
or ``most natural''	such ensemble.	
	
We quantify	the	{\em robustness} of	a particular state 
$\ket{\psi_k^{\cal U}}$	by its survival	probability	$S_k^{\cal U}(t)$. 
This	is the
probability	that the system	would be found (by a hypothetical
projective measurement)	to still be	in the state $\ket{\psi_k^{\cal U}}$
at time	$t$. It	is given by	
\beq   \label{overlap}
S_k^{\cal U}(t)	= \bra{\psi_k^{\cal U}}	e^{{\cal L}t}
\sq{\ket{\psi_k^{\cal U}}\bra{\psi_k^{\cal U}}} \ket{\psi_k^{\cal U}}.
\eeq
	
Since we are considering an	ensemble $E^{\cal U}$ we must define	
the	average	survival probability	
\beq \label{defsp}
S^{\cal	U}(t) =	\sum_{k} \wp_k^{\cal U} S_k^{\cal U}(t).
\eeq
In the limit $t\to\infty$ the ensemble-averaged	survival	
probability	will tend towards the stationary value	
\beq
S^{\cal	U}(\infty) = {\rm Tr}[ \rho_{\rm ss} ^2].	
\eeq
This is	independent	of the unraveling ${\cal U}$ and is	a measure of
the	mixedness of $\rho_{\rm	ss}$.

There are many possible figures-of-merit that may be obtained from 
the survival probability $S^{\cal U}(t)$, as discussed in 
Ref.~\cite{WisVac00}. Here we choose the simplest one, also adopted in 
\cite{WisVac00}: 
the	time it	takes for $S^{\cal U}(t)$ to fall half-way to its equilibrium 
value. That is,
\beq \label{deftau}	
\tau^{\cal U} =	{\rm min} \{ t\,:\,	S^{\cal	U}(t)= (1+{\rm 
Tr}[\rho_{\rm ss}^{2}])/2	\}.
\eeq
This {\em survival time} $\tau^{\cal U}$ quantifies the robustness	of 
a particular 	
unraveling ${\cal U}$.

Let	the	set	of all unravelings be denoted $J$.
Then the subset	of {\em	maximally robust} unravelings $J_{M}$ is	
\beq
J_{M} =	\{ {\cal R}	\in	J :	\tau^{\cal R} \geq \tau^{\cal U} \;	\forall
\, {\cal	
U} \in J \}.
\eeq
Even if	$J_{M}$	has	many elements ${\cal R}_{1}, {\cal R}_{2}, \ldots$,
these different	unravelings	may
give the same ensemble $E^{\cal	R}=E^{{\cal	R}_{1}}=E^{{\cal
R}_{2}}=\ldots$.	
In this	case we claim $E^{\cal R}$ is
the	most natural ensemble representation of	the
stationary solution	of a given master equation.
	Different definitions of survival time \cite{WisVac98,WisVac00} will 
obviously lead to different numerical values for $\tau^{\cal R}$. We 
are less concerned with such numerical values than with the robust 
ensemble $E^{\cal R}$, which has been found \cite{WisVac00} to 
depend little on the precise definition 
used.

\section{The Two-Level Atom}

\subsection{The Resonance Fluorescence Master Equation}

Consider an atom with two relevant levels $\{\ket{g},\ket{e}\}$. 
Let  there be a dipole moment between these levels so that the coupling to 
the continuum of electromagnetic field modes in the vacuum state will 
cause the atom to decay  at 
rate $\gamma$. So that the atom does not simply decay to the state 
$\ket{g}$, add driving by a classical field (such as that produced by 
a laser) of Rabi frequency $\Omega$. 
We work in the interaction picture with respect to the free 
Hamiltonian $H_{0}=\hbar\omega_{0}\ket{e}\bra{e}$ so that the 
classical driving at frequency $\omega$ becomes time-independent. 
The evolution of the atom's state matrix can then be 
described by the resonance fluorescence (RF) master equation 
\beq \label{me2}
\dot{\rho} = -i\frac{\Omega}{2}[\sigma_{x},\rho] + \gamma {\cal D}[\sigma]\rho.
\eeq

In this equation we have used the  Pauli matrices
\bqa 
\sigma_x & = & | e\rangle \langle g| +| g\rangle \langle e| \\
\sigma_y & = & -i| e\rangle \langle g| +i| g\rangle \langle e| \\
\sigma_z & = & | e\rangle \langle e| -| g\rangle \langle g| \\
\sigma & = & | g\rangle \langle e|
=\frac 12(\sigma_x -i\sigma_y )\\ 
\sigma^{\dagger }& = & | e\rangle \langle g| 
=\frac 12(\sigma_x +i\sigma_y )
\eqa
In terms of these, any state of the atom can be written
as a 3-vector $(x,y,z)$ satisfying
\beq
x^{2}+y^{2}+z^{2}\leq 1,
\eeq
with equality only for a pure state. From this Bloch vector the state 
matrix is defined by
\beq
\rho = \frac 12\left( I+x\sigma_x 
+y\sigma_y +z\sigma_z \right) .
\eeq

The linear equations of motion for the Bloch vector that result
from \erf{me2} are known as the Bloch equations. The solution 
satisfying the initial condition
\beq
x(0)=u\;\;,\;y(0)=v\;\;,\;z(0)=w
\eeq
is
\bqa 
x(t) & = & ue^{-(\gamma/2)t}, \label{x} \\ 
y(t) & = & c_{+}e^{\lambda_+ t}+c_{-}e^{\lambda_- t}+y_{\rm ss},\\ 
z(t) & = & c_{+}\frac {\gamma-4i\tilde\Omega}{4\Omega } 
e^{\lambda_+ t}+c_{-}\frac {\gamma+4i\tilde\Omega}{4\Omega }
 e^{\lambda_- t}+z_{\rm ss} .\label{z}
\eqa
Here $c_{\pm}$ are constants given by
\beq
c_{\pm} = \frac{1}{8i\tilde \Omega }\left[\mp 4\Omega 
(w-z_{\rm ss} ) \pm
(\gamma \pm 4i\tilde \Omega )(v-y_{\rm ss})\right] .
\eeq
The eigenvalues $\lambda_{\pm}$ are defined by
\beq
\lambda_{\pm} = -\frac{3}{4}\gamma\pm i\tilde \Omega.
\eeq
Here
\beq \label{deftom}
\tilde{\Omega} = \sqrt {\Omega^2 -({\gamma }/{4})^2}
\eeq
is a real modified Rabi frequency for $\Omega > \gamma/4$, and is
imaginary for $\Omega < \gamma/4$.
The stationary solutions appearing in the above equations are
\bqa
 x_{\rm ss} & = & 0,\\ 
 y_{\rm ss} & = & \frac {2\gamma \Omega }{\gamma^2 +2\Omega^2 },\\ 
 z_{\rm ss} & = & \frac 
{-\gamma^2 }{\gamma^2 +2\Omega^2} .
\eqa

\subsection{The Survival Probability}

Using the Bloch vector representation of the atomic state matrix it is 
easy to show that the survival probability for a pure state $\ket{\psi_k}$ 
with projector 
\beq
\ket{\psi_k}\bra{\psi_k} = \frac12  \ro{1+u_k\sigma_x + v_k\sigma_y+ 
w_k\sigma_z}
\eeq
is  
\beq
S_k(t) = \frac12 
\ro{1+x_k(t)u_k + y_k(t)v_k +z_k(t)w_k },
\eeq
where $(x,y,z)_{k}(t)$ is the Bloch vector at time $t$ with the initial 
condition $(x,y,z)_{k}(0)=(u,v,w)_{k}$ as in 
\erfs{x}{z}. From that solution it is evident that $S_{k}(t)$ will 
contain terms that are constant, linear, and bilinear in the vector 
components $(u,v,w)_{k}$ of the initial state.

As explained in the preceding section we are interested in the survival 
probability not for a single state  but for an ensemble of states. This is 
the ensemble average of \erf{defsp}. After some work, the 
survival probability in this case is found to be simply
\bqa
S(t) &=& \sum_{k} \wp_{k}S_{k}(t) \\
&=& \frac12 (1+y_{\rm ss}^{2} + z_{\rm 
ss}^{2}) + \frac12 \label{sp}\\
&&\times\; \sq{(1-y_{\rm ss}^{2} - z_{\rm 
ss}^{2})e^{-(\gamma /2)t}+ V_{v} f_{+}(t) + V_{w}f_{-}(t)}, \nn
\eqa
where
\beq \label{deffpm}
f_{\pm}(t) = -e^{-(\gamma/2)t} + e^{-(3\gamma /4)t}\ro{\cos\tilde\Omega t
\pm \frac{\gamma}{4\tilde\Omega}\sin\tilde\Omega t}.
\eeq
In \erf{sp}, all the information about the ensemble is contained in 
the moments
\bqa
V_{v} &\equiv& {\rm E}[v^2] - E[v]^{2}=\sum_{k} \wp_k v_k^2 - 
\ro{\sum_{k}\wp_{k}v_{k}}^{2}, \\
V_{w} &\equiv& {\rm E}[w^2] - E[w]^{2}=\sum_{k} \wp_k w_k^2 - 
\ro{\sum_{k}\wp_{k}w_{k}}^{2}.
\eqa
This is possible because we have used the following relations: 
\bqa
{\rm E}[(u,v,w)] &=& (x_{\rm ss},y_{\rm ss},z_{\rm ss}) ,\\
{\rm E}[u^{2}] &=& 1-{\rm E}[v^{2}] - {\rm E}[w^{2}].
\eqa
To find the robustness of any 
particular unraveling we thus need to find simply the two ensemble 
averages $V_{v}$ and $V_{w}$.

\section{Unraveling by Direct Detection}

The most obvious way to unravel the RF master equation is by direct 
detection. This requires detecting all of the atom's fluorescence by 
unit-efficiency photodetectors. This is beyond current technology, but not 
by so much that the experiment should be considered unphysical. As we will 
find, unraveling by direct detection  is actually not very  robust (by the 
definition of Sec.~II)  but it is nevertheless useful to consider as a 
point of comparison with more robust unravelings.

The stochastic evolution of an atom undergoing RF with direct 
detection has been considered many 
times before \cite{Car93b,WisMil93c}. 
It has one feature that enables an enormous 
simplification over a generic unraveling. This is that immediately 
following a detection, the atomic state is independent of its state 
before, and is just the ground state $\ket{g}$. Between these 
jumps to the ground state, the conditioned atomic state evolves 
deterministically. At steady state, when there has certainly been at 
least one  detection, all members of the ensemble are therefore identified 
simply by the time $t$ since the last detection.

 If there is a detection at time $t_0$, then, until the next detection 
 occurs,  the state of the atom at time $t_{0}+t$ evolves 
according to the equation \cite{Car93b,WisMil93c}
\beq \label{modrabi}
\frac d{dt}| \tilde 
\psi_{0} (t)\rangle = -\left( \frac {\gamma 
}2\sigma^{\dagger }\sigma -i\frac{\Omega}{2}\sigma_{x}\right) 
| \tilde \psi_{0} (
t)\rangle .
\eeq
The solution, satisfying the initial condition 
$\ket{\tilde\psi_{0}(0)} = \ket{g}$ is 
\beq \label{unnorm}
|\tilde \psi_0 \rangle =\tilde c_e(t)| e\rangle 
+\tilde c_g(t)| g\rangle ,
\eeq
where
\bqa
\tilde{c}_{g}(t) &=& \sq{\cos(\check{\Omega}t/2) + 
\frac{\gamma}{2\check\Omega}\sin(\check\Omega t/2) }e^{-(\gamma/4)t},\\
\tilde{c}_{e}(t) &=& -i\frac{\Omega}{\check{\Omega}} \sin(\check\Omega 
t/2)e^{-(\gamma/4)t}.
\eqa
Here 
\beq
\check\Omega = \sqrt{\Omega^{2}-(\gamma/2)^{2}}
\eeq
 is a real modified Rabi frequency for $\Omega > \gamma/2$ and is 
 imaginary for $\Omega<\gamma/2$. Note that it is different from 
 $\tilde\Omega$ defined in \erf{deftom}.

The state in \erf{unnorm} is unnormalized, and the norm
$\ip{\tilde\psi_{0}(t)}{\tilde\psi_{0}(t)}$
represents the probability that there has been no detection 
since time $t_0$, given that there was a detection at that time.
Let us write this probability as
\beq \label{defP0}
P_{0}(t) = |\tilde{c}_{e}(t)|^{2} + |\tilde{c}_{g}(t)|^{2}.
\eeq
We show in the appendix that this probability is related to 
$\wp(t)$, the probability that, at steady state, the last detection 
was a time $t$ ago, by
\beq \label{defwp}
\wp(t) = \frac{P_{0}(t)}{\int_{0}^{\infty}P_{0}(s)ds}.
\eeq

As noted above, in steady state under direct detection the possible 
atomic states are parametrized by the real variable $t$, the time 
since the last detection. The state at that time has projector
\beq \label{ddhatP}
\hat P(t) = \frac{\ket{\tilde\psi_{0}(t)}\bra{\tilde\psi_{0}(t)}}
{\ip{\tilde\psi_{0}(t)}{\tilde\psi_{0}(t)}},
\eeq
and the weight for each of these members of the ensemble is $\wp(t)dt$. 
Physically, all members of the ensemble exist on the $u=0$ great 
circle of the Bloch sphere, because that is where the modified Rabi 
cycling of \erf{modrabi} takes the ground state. 
This distribution is shown in Fig.~\ref{fig:bs}(a). For $\check\Omega$ 
imaginary (that is, $\Omega < 
\gamma/2$) the states never reach the excited state. For 
$\check\Omega$ real (that is, $\Omega > \gamma/2$, the states may 
undergo an arbitrary number of cycles. For $\Omega \gg \gamma$ the 
states are likely to undergo many cycles before a spontaneous 
emission event occurs so that the ensemble consists of all the states 
on the $u=0$ great circle, almost uniformly distributed.

From \erf{ddhatP} it can be verified analytically that
\beq
\int_{0}^{\infty} \hat{P}(t) \wp(t) dt = \frac{1}{\cal N}\int_{0}^{\infty}dt 
\ket{\tilde\psi_{0}(t)}\bra{\tilde\psi_{0}(t)} = \rho_{\rm ss},
\eeq
where
\beq
{\cal N} = \int_{0}^{\infty} \sq{|\tilde{c}_{e}(t)|^{2}
+ |\tilde{c}_{g}(t)|^{2}} dt.
\eeq.
Moreover we can easily find numerically the ensemble averages 
necessary to find the ensemble average survival probability, namely
\bqa
V_{v} & = &\frac{1}{\cal N} 
\int_{0}^{\infty} 
\frac{[i\tilde c_{e}(t)\tilde c_{g}^{*}(t)-i\tilde c_{e}^{*}(t)\tilde 
c_{g}(t)]^{2}}{|\tilde{c}_{e}(t)|^{2}
+ |\tilde{c}_{g}(t)|^{2}}dt - y_{\rm ss}^{2},\\
V_{w} & = & \frac{1}{\cal N}
\int_{0}^{\infty} 
\frac{[|\tilde{c}_{e}(t)|^{2}
- |\tilde{c}_{g}(t)|^{2}]^{2}}
{|\tilde{c}_{e}(t)|^{2}
+ |\tilde{c}_{g}(t)|^{2}}dt - z_{\rm ss}^{2},
\eqa

\begin{figure}
\includegraphics[width=0.45\textwidth]{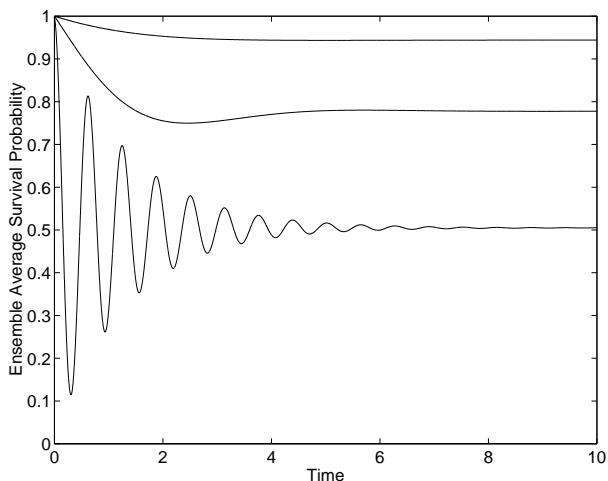}
\caption{\narrowtext Plot of the ensemble average survival 
probability $S(t)$ versus time (in units of $\gamma^{-1}$) for the 
stationary ensemble of direct detection. The three curves are, from 
the top down, $\Omega = \gamma/2$, $\Omega=\gamma$, and  
$\Omega=10\gamma$.}
	\protect\label{fig:sp}
\end{figure}

In Fig.~\ref{fig:sp} we plot the survival probability for the direct detection 
ensemble, for a variety of driving strengths $\Omega$. We see that 
for $\Omega < \gamma$ the survival probability decays approximately 
exponentially at rate of order $\gamma$. For $\Omega \ll \gamma$ the 
stationary state matrix is close to the ground state, and most members 
of the direct detection ensemble are also. 
For $\Omega \gg \gamma$ the ensemble is 
equally spread over the $u=0$ great circle and consequently the 
variances $V_{v}$ and $V_{w}$ are approximately equal to $1/2$. Using 
this, the survival probability is found to be 
approximately
\beq \label{spapp}
S(t) \approx \frac12 \ro{1+e^{-(3/4)\gamma t}\cos\tilde\Omega t}.
\eeq

The oscillations in the survival probability are 
due to the Rabi oscillations. As noted 
above, the direct detection ensemble consists of states on the $u=0$ 
great circle. Rabi cycling around the $x$-axis according to the RF 
master equation (\ref{me2}) rotates this circle 
around, rapidly moving the states away from their initial positions 
and then back close to their initial conditions after one cycle. They 
do not return exactly to their initial states because of the slow (at 
rate $3\gamma/4$) decay towards the equilibrium state. This 
behaviour is illustrated for a typical member of the direct detection 
ensemble in Fig.~\ref{fig:bs}(b).

\begin{figure}
\includegraphics[width=0.45\textwidth]{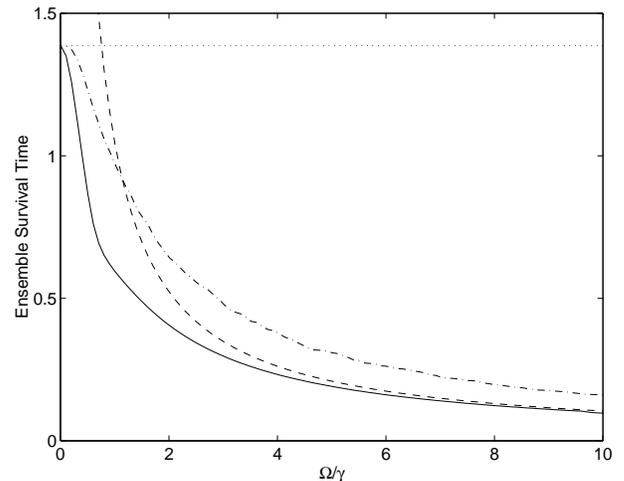}
	\caption{\narrowtext Plot of the ensemble survival time (in units of 
	$\gamma^{-1}$) versus $\Omega/\gamma$. The solid curve is for the 
	ensemble arising from direct detection (Sec.~IV). The dashed curve is the 
	large $\Omega/\gamma$ analytical approximation to it in \erf{anapp}. 
	The dotted curve is that from 
	the maximally robust unraveling (Sec.~V). 
	The dash-dot curve is that from the 
	maximally robust continuous Markovian unraveling (Sec.~VI). The 
	unevenness in this final curve is due to statistical error 
	in the ensemble averages $V_{v}$ and $V_{w}$), and gives an 
	indication of the size of the statistical error. }
	\protect\label{fig:st}
\end{figure}

The change from damped to oscillatory behaviour has a dramatic effect 
on the survival time in \erf{deftau}. It is plotted in 
Fig.~\ref{fig:st} as 
a function of $\Omega$. For $\Omega \ll \gamma$ it is
given by
\beq
\tau \simeq 2\ln 2 \gamma^{-1},
\eeq
as in this limit the survival probability decays 
as $e^{-\gamma t/2}$. For $\Omega \gg \gamma$, we can use
\erf{spapp} to get the approximate 
expression 
\beq \label{anapp}
\tau \simeq \frac{\pi}{3}\Omega^{-1}.
\eeq
That is, the survival time is here determined by the Hamiltonian 
evolution only.
As shown in Fig.~\ref{fig:st}, this is quite a good approximation even for 
moderate $\Omega$.

\section{The Most Robust Unraveling}
\label{sec:mru}

\subsection{The Most Robust Ensemble}

It was stated above that the unraveling by direct detection is not the 
most robust unraveling. In fact, from an examination of the survival 
probability in \erf{sp} 
 we can see that it is one of the 
least robust unravelings for $\Omega \gg \gamma$. 
That is because of the large variances in 
$v$ and $w$ in this limit. 

It is not difficult to show that the two 
functions $f_{+}(t)$ and $f_{-}(t)$, defined in \erf{deffpm}, 
are non-positive for all $\Omega$ 
and for $t>0$. Since the variances $V_{v}$ and $V_{w}$ are 
non-negative, it is easy to see that to maximize the survival 
probability, one would wish to minimize $V_{v}$ and $V_{w}$. The ideal 
limit would be $V_{v}=V_{w}=0$. This corresponds to an ensemble in 
which all members have the same Bloch vector components $v$ and $w$. 
Since the ensemble average must equal $\rho_{\rm ss}$ it follows then 
that for all members
\beq
v=y_{\rm ss}\;,\;\;w=z_{\rm ss}.
\eeq
Furthermore, since $x_{\rm ss}=0$, and since the members of the 
ensemble must be pure, it follows that for all members
\beq
u = \pm \sqrt{1-y_{\rm ss}^{2} - z_{\rm ss}^{2}},
\eeq
where the two alternatives are equally weighted.
These two members of the ensemble are shown in Fig.~\ref{fig:bs}(c).

\begin{figure}
\includegraphics[width=0.45\textwidth]{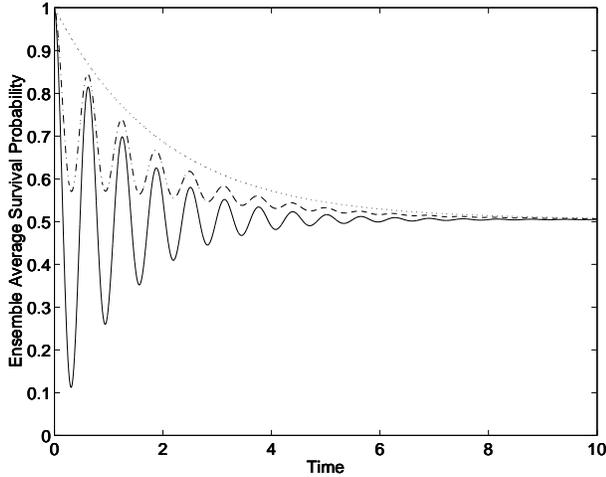}
\caption{\narrowtext Plot of the ensemble average survival 
probability $S(t)$ versus time (in units of $\gamma^{-1}$) for the 
stationary ensembles of various detection schemes all with 
$\Omega/\gamma=10$. The three curves are for 
the minimally robust ($\upsilon=-1$) continuous Markovian unraveling (solid), 
the maximally robust ($\upsilon=1$) continuous Markovian 
unraveling (dash-dot) and the maximally robust unraveling (dotted).}
	\protect\label{fig:sp2}
\end{figure}

This ensemble is guaranteed to give the 
maximum survival probability
\beq
S(t) = \frac12 (1+y_{\rm ss}^{2} + z_{\rm 
ss}^{2}) + \frac{1}{2}\sq{(1-y_{\rm ss}^{2} - z_{\rm 
ss}^{2})e^{-(\gamma /2)t}},
\eeq
which is plotted in Fig.~\ref{fig:sp2}. It gives the maximum survival time
\beq
\tau = 2\ln 2 \gamma^{-1},
\eeq
which is independent of $\Omega$. There is no Rabi cycling because the states 
defined by the Bloch vector $(u,v,w)$ have $v$ and $w$ already equal 
to their stationary values. Under the master equation evolution $u$ 
simply decays towards its stationary value of zero and $v$ and $w$ 
remain constant. This decay of the Bloch vector 
to equilibrium at rate $\gamma/2$ 
is shown in Fig.~\ref{fig:bs}(d).

\subsection{Adaptive Interferometric Detection}

\begin{figure}
\includegraphics[width=0.45\textwidth]{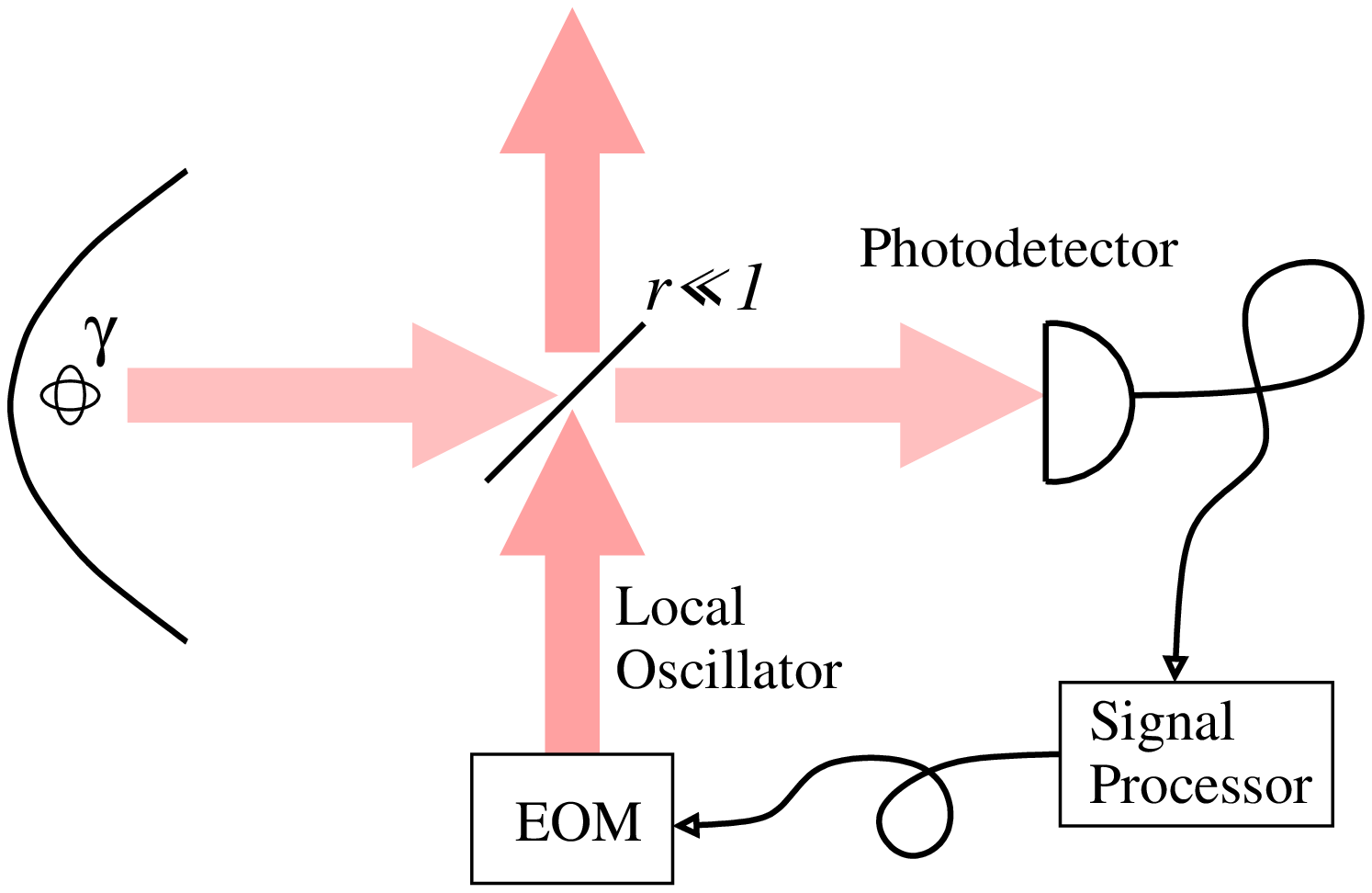}
	\caption{\narrowtext  Diagram of the experimental configuration 
	for an adaptive interferometric measurement of the fluorescence of an atom.
	The signal and a local oscillator are incident on a beam splitter of 
	reflectance $r \ll 1$. The 
	amplitude of the local oscillator is variable as a 
	function of time, determined by an electro-optic modulator (EOM).
	The modulator is controlled by the experimenter, being inverted 
	every time a detection occurs.}
	\protect\label{fig:adapt}
\end{figure}

The most robust ensemble defined here would be a mere curiosity if it 
were not for the fact that there is a detection scheme that realizes 
it. This scheme, proposed by one of us and Toombes \cite{WisToo99}, 
involves interfering the light from the atom 
with a resonant local oscillator before detection. This is done using a
 highly transmitting 
beam splitter as shown in Fig.~\ref{fig:adapt}. In the limit of a large local 
oscillator, this is known as homodyne detection. However we require a 
very weak local oscillator, with reflected intensity comparable to the 
intensity of the light from the atom. Furthermore, we require the 
local oscillator amplitude to be continually adjusted by a real-time 
feedback loop. To be specific, the field detected should be 
proportional to 
\beq
\sigma + \mu(t),
\eeq
where the field from the atom is proportional to 
$\sigma$, the atomic lowering operator as usual, and the local 
oscillator field is represented by the complex number $\mu(t)$. This 
complex number is given by
\beq
\mu(t) = \pm \frac12,
\eeq
where the sign is changed every time a detection occurs.

Remarkably, this relatively simple detection scheme has the 
consequence that, after initial transients have died away, 
a driven atom jumps between the states with projector $\hat{P}_{\pm}$, where
\beq \label{ppm}
\hat{P}_{\pm} = \frac12 \ro{I \pm \sqrt{1-y_{\rm ss}^{2} - z_{\rm ss}^{2}}
\sigma_x + v_{\rm ss}\sigma_y+ 
w_{\rm ss}\sigma_z},
\eeq
every time a detection occurs. The rate of detections moreover is 
independent of which of these states the atom is in, and is equal 
to $\gamma/4$. Thus in the long time limit the atom will have a 
probability $\wp_{\pm}=1/2$ to be in each of them, and the maximally 
robust ensemble will be physically realized.

\section{Continuous Markovian Unravelings}

\label{sec:mrcmu}

The most robust unraveling is neither continuous nor 
Markovian. It is not continuous because the atomic state jumps every 
time there is a detection. It is not Markovian because the evolution 
of the atomic state does not depend only on its present state. Rather, 
it depends on the past history of detections through the local 
oscillator amplitude $\mu(t)$. 
As noted in the introduction, previous investigations of robust 
unravelings in other systems\cite{WisVac98,WisVac00} have 
been restricted to unravelings that do have these properties. It is therefore 
interesting to ask for the present system, 
how close to the MRU is the most robust unraveling 
that is continuous and Markovian?

Continuous Markovian unraveling (CMU) 
of the RF master equation (\ref{me2}) can be 
represented by a two-parameter family of nonlinear SSEs 
for the non-normalized state vector $\ket{\bar\psi(t)}$ of the form
\beq \label{cmu1}
d\ket{\bar{\psi}(t)}=dt\sq{-iH - 
\frac{\gamma}{2}\sigma\dg \sigma
+ J(t) \sigma }\ket{\bar{\psi}(t)} ,
\eeq
which is to be interpreted in the \ito sense \cite{Gar85}. Here 
$J(t)$ is a complex ``current'' given by
\beq
J(t)dt = \gamma\an{\upsilon \sigma+\sigma\dg}dt + \rt\gamma dW(t),
\eeq
where $\upsilon $ is a complex number satisfying
\beq
\upsilon ^*\upsilon  \leq 1,
\eeq
and the angle brackets denote a quantum average using the normalized 
state vector $\ket{\psi(t)}$. We use $\ket{\bar\psi}$ rather than 
$\ket{\tilde\psi}$ because the norm of $\ket{\bar\psi}$ has no 
interpretation in terms of probability, unlike that of 
$\ket{\tilde\psi}$ in Sec.~IV. The stochastic term $dW(t)$ 
 is a  complex Gaussian white noise term  
satisfying
\bqa
{\rm E}[dW] &=& 0, \label{mean0} \\
{\rm E}[dW^*dW] &=& dt ,\label{vardt} \\
{\rm E}[(dW)^2] &=& \upsilon dt.
\eqa

The complex parameter $\upsilon $ 
comprises the two parameters for the family of unravelings of the 
form of \erf{cmu1}. From \erf{mean0} and \erf{vardt} 
it can be shown that \erf{ensav} is satisfied for all $\upsilon$.  
Thus 
\beq
\rho(t) = {\rm E}\sq{\frac{\ket{\bar{\psi}(t)}\bra{\bar{\psi}(t)}}
{\ip{\bar{\psi}(t)}{\bar{\psi}(t)}}} 
\eeq
obeys the RF master equation (\ref{me2}), while in an individual 
trajectory the state remains pure. In this case there is no simple 
way to find the steady state ensemble of pure states. A numerical 
simulation of \erf{cmu1}, with time averages replacing ensemble 
averages, is the only way to proceed.

Finding the maximally robust continuous Markovian unraveling (MRCMU) in this case 
reduces to a search over the ball $|\upsilon |^{2} \leq 1$ in the complex plane.
We find that the MRCMU is for $\upsilon =1$. For this value of $\upsilon$ the 
``current'' $J(t)$ (which is real in this case) has a deterministic part 
equal to 
\beq
{\rm E}\sq{J(t)} = \gamma\an{\sigma_{x}}.
\eeq
That is, the measurement yields information about the $\sigma_{x}$ 
quadrature of the atomic dipole. As a consequence it 
tends to localize the atom near the $\sigma_{x}$ eigenstates. This 
localization is relatively stable for high driving, since 
these are also eigenstates of the Hamiltonian $\Omega \sigma_{x}/2$. 
This is shown in Fig.~\ref{fig:bs}(e), which is a 
stochastically generated sample of 10000 states from the equilibrium 
ensemble for the MRCMU. In complete contrast to the direct detection ensemble 
in Fig.~\ref{fig:bs}(a), most of the states lie close to the 
$\sigma_{x}$ eigenstates. For $\Omega$ large as in this figure, this
means close to the two members of the MRU shown in 
Fig.~\ref{fig:bs}(c). 

Because a typical member of the MRCMU ensemble is fairly close to 
the $\sigma_{x}$ axis, it is relatively little affected by the 
Hamiltonian, which causes rotation around that axis. Under the RF 
master equation (\ref{me2}) its evolution consists of decay to the 
equilibrium, with relatively small 
Rabi oscillations superimposed. This is as shown in 
Fig.~\ref{fig:bs}(f). As a consequence, although the survival probability 
oscillates, it remains above its equilibrium value and decays 
towards it at a rate proportional to $\gamma$. This is  
shown in Fig.~\ref{fig:sp2} for $\Omega=10\gamma$. Also shown, for 
comparison, is the survival probability for the minimally robust 
CMU, which occurs for $\upsilon=-1$. This is 
almost identical to the corresponding curve for direct detection in 
Fig.~\ref{fig:sp}, as the ensemble is confined to the $u=0$ great 
circle in both cases.

The survival time for the MRCMU is shown in Fig.~\ref{fig:st}. 
Because of our definition of survival time in \erf{deftau}, the 
survival time for the MRCMU is determined by the rapid (for $\Omega 
\gg \gamma$) oscillations in the 
survival probability rather than the slow mean decay.
Thus it is qualitatively similar to the survival time 
for the direct detection ensemble,  starting at $2\ln 2\gamma^{-1}$ and then 
falling like $\Omega^{-1}$ as $\Omega$ increases. 

\section{Comparison of the MRU and the MRCMU}

We have seen that no continuous Markovian unraveling (CMU) is as robust as 
the maximally robust unraveling (MRU), which is neither continuous nor 
Markovian. Furthermore, the robustness for the maximally robust CMU, 
as measured by the survival time, scales in the same way as that for 
the very non-robust direct detection scheme. However, because of the
 arbitrariness in any definition of survival time, 
we are interested more in the robust ensembles themselves than in the numerical 
values of their survival times. As discussed above, the 
 MRCMU realizes an ensemble that has a common 
feature with that realized by the MRU: the states in the 
ensemble tend to have well-defined values of $u=\an{\sigma_{x}}$. 
In this section we wish to answer the question: just how close is the 
MRCMU ensemble to the MRU ensemble?

\subsection{Closeness of Two Ensembles}

To answer this question we require a measure (not necessarily 
transitive) from ensemble $E^{A}$ to ensemble $E^{B}$, where each 
ensemble represents
the same state matrix. 
It seems best to choose an operationally defined 
measure, which we do as follows. 

By allowing the same projector to 
reappear with different indices, we can write, to any desired degree 
of accuracy,
\bqa
E^{A}&=&\cu{(\ket{\phi_{k}},N^{-1}):1\leq k \leq N},\\
E^{B}&=&\cu{(\ket{\psi_{\mu}},N^{-1}):1\leq \mu \leq N},
\eqa
where $\ket{\phi_{k}}$ and $\ket{\psi_{\mu}}$ are normalized states
 such that
\beq
\rho = N^{-1}\sum_{k}\proj{\phi_{k}}
 = N^{-1}\sum_{\mu}\proj{\psi_{\mu}},
\eeq
where $N$ is an arbitrarily large integer.

To define the closeness of the ensembles, imagine that there are two 
people, Alice and Bob. Alice has in her possession a measuring device 
with $N$ settings, corresponding to the $N$ projectors 
$\{\proj{\phi_{k}}:k\}$. If setting $k$ is chosen then the device makes a 
projective measurement with projector $\proj{\phi_{k}}$. Bob has in his 
possession an ensemble of quantum states $\{\ket{\psi_{\mu}}:\mu\}$. It is 
Bob's aim to try to convince Alice that he is actually in possession 
of the ensemble $\{\ket{\phi_{k}}:k\}$. He must submit each of his $N$ 
systems $\ket{\psi_{\mu}}$ to Alice, telling her which of her states 
$\ket{\phi_{k}}$ 
each is supposed to be in. She then makes the appropriate measurement 
and, unless Bob's ensemble really is the same as Alice's, is likely 
to find errors some of the time. An error is when a state that Bob claims is 
$\ket{\psi_{k}}$ is found to give the answer ``no'' to Alice's projective 
measurement ``is the state $\ket{\phi_{k}}$?'' Assuming Bob chooses a good 
strategy, then the higher the probability 
of error, the larger the distance between the two ensembles. 

We can formalize this as follows. Say Bob actually sends state 
$\ket{\psi_{\mu}}$, but claims it is $\ket{\phi_{k(\mu)}}$, where the 
functional dependence here indicates that Bob makes his choice of 
index $k$ based on his actual state.  Then the probability 
of error for this state is
\beq
\epsilon_{k(\mu)|\mu} = 1-\left|\ip{\phi_{k(\mu)}}{\psi_{\mu}}\right|^{2}.
\eeq
The ensemble average probability of error, for Bob's optimum strategy, is
\beq
\epsilon_{\rm opt} = N^{-1} \min_{M} \sum_{\mu=1}^{N} \epsilon_{k(\mu)|\mu},
\eeq
where the minimum is over all one-to-one mappings $M$
\beq
\mu \stackrel{M}{\longrightarrow} k(\mu)\;,\;\;
k \stackrel{M^{-1}}{\longrightarrow} \mu(k).
\eeq
Bob's strategies have to correspond to a mapping of this form because 
unless Bob names each of Alice's states once and once only Alice 
would know that Bob is lying when he claims to be in possession of 
Alice's ensemble $\{\ket{\phi_{k}}:k\}$.

We could take the distance between the ensembles to be equal to this 
minimum average error probability. However, if the state matrix $\rho$ is 
close to being pure (with ${\rm Tr}[\rho^{2}]$ close to one),
 then the average error probability would be 
small regardless of what strategy Bob chose. In particular, if Bob 
had a totally random strategy then the error probability for an 
individual state $\ket{\psi_{\mu}}$ would be, on average,
\beq
\langle\epsilon_{k|\mu}\rangle = 
1-\bra{\psi_{\mu}}\overline{\proj{\phi_{k}}}\ket{\psi_{\mu}}
= 1-\bra{\psi_{\mu}}\rho \ket{\psi_{\mu}} ,
\eeq
and the ensemble average error probability would be
\beq
\overline{\epsilon} = N^{-1} \sum_{\mu=1}^{N} \ro{ 1-{\rm Tr}[\rho 
\proj{\psi_{\mu}}]} = 1-{\rm Tr}[\rho^{2}].
\eeq
That is, the average error probability would be close to zero even 
though Bob does not take into account the difference between his 
states so that his effective ensemble consists of $N$ copies of the 
mixed state $\rho$.

For this reason, it seems better to define the distance between the 
ensembles by the normalized error probability:
\beq
d(E^{A}|E^{B}) = \frac{1 - \min_{M} 
\sum_{\mu=1}^{N} \frac{1}{N}\left|\ip{\phi_{k(\mu)}}{\psi_{\mu}}\right|^{2}}
{1-{\rm Tr}[\rho^{2}]}.
\eeq
In this case it is easy (at least for a two-level system) to see that
\beq
0\leq d(E^{B}|E^{A}) \leq 1,
\eeq
where the lower bound is attained if and only if the ensembles are 
identical, 
and where no tighter upper bound can be found for a given $\rho$. 
Thus the two ensembles could be said to be close if and only if 
$d(E^{B}|E^{A})\ll 1$.

\subsection{Closeness of the MRU and MRCMU Ensembles}

In the case at hand the reference ensemble $E^{A}$ is the most 
robust ensemble of Sec.~\ref{sec:mru}, while the ensemble $E^{B}$ 
whose closeness we 
wish to gauge is that of the most robust continuous Markovian unraveling 
of Sec.~\ref{sec:mrcmu}. The fact that $E^{A}$ has only two elements, 
whereas $E^{B}$ has an infinitude of elements causes no problems. 
We simply allow an arbitrarily large number $N$ of elements for each 
ensemble and let half of those for ensemble $A$ be the state with 
projector $\hat P_{+}$ 
and half the state $\hat{P}_{-}$, as defined in \erf{ppm}. 

Because the two states $\hat{P}_{\pm}$ differ 
only by the sign of $\an{\sigma_{x}}$, and because both ensembles are 
symmetric about a reflection in the $y-z$ plane, 
Bob's best strategy is easy to 
find. For each of his states $\ket{\psi_{\mu}}$ he tells Alice that it is 
state $\hat{P}_{+}$ if it has a positive mean $\sigma_{x}$, and $\hat 
P_{-}$ if it has a negative mean $\sigma_{x}$. If $\ket{\psi_{\mu}}$ is 
the state
\beq
\proj{\psi_{\mu}} = \frac{1}{2}(1+ u_{\mu}\sigma_{x}+v_{\mu}\sigma_{y}
+w_{\mu}\sigma_{z}),
\eeq
then the probability of
error for this state is
\bqa
\epsilon_{\mu} &=& 1-\bra{\psi_{\mu}} \hat P_{{\rm 
sign}(u_{\mu})}\ket{\psi_{\mu}}\\
&=& \frac12 - \frac12(|u_{\mu}|\sqrt{1-y_{\rm ss}^{2} - z_{\rm 
ss}^{2}}+v_{\mu}y_{\rm ss}+w_{\mu} z_{\rm ss}).
\eqa
Averaging over all of Bob's states we get
\beq
\epsilon_{\rm opt} = \frac12 - \frac12\ro{\frac{1}{N}\sum_{\mu}|u_{\mu}|\sqrt{1-y_{\rm ss}^{2} - z_{\rm 
ss}^{2}}+y^{2}_{\rm ss}+ z^{2}_{\rm ss}},
\eeq
and the distance from the MRU ensemble to the MRCMU ensemble is
\beq
d(E^{A}|E^{B}) = 1 - \frac{{\rm E}[|u|]}{\sqrt{1-y_{\rm ss}^{2}-z_{\rm 
ss}^{2}}}.
\eeq

\begin{figure}
\includegraphics[width=0.45\textwidth]{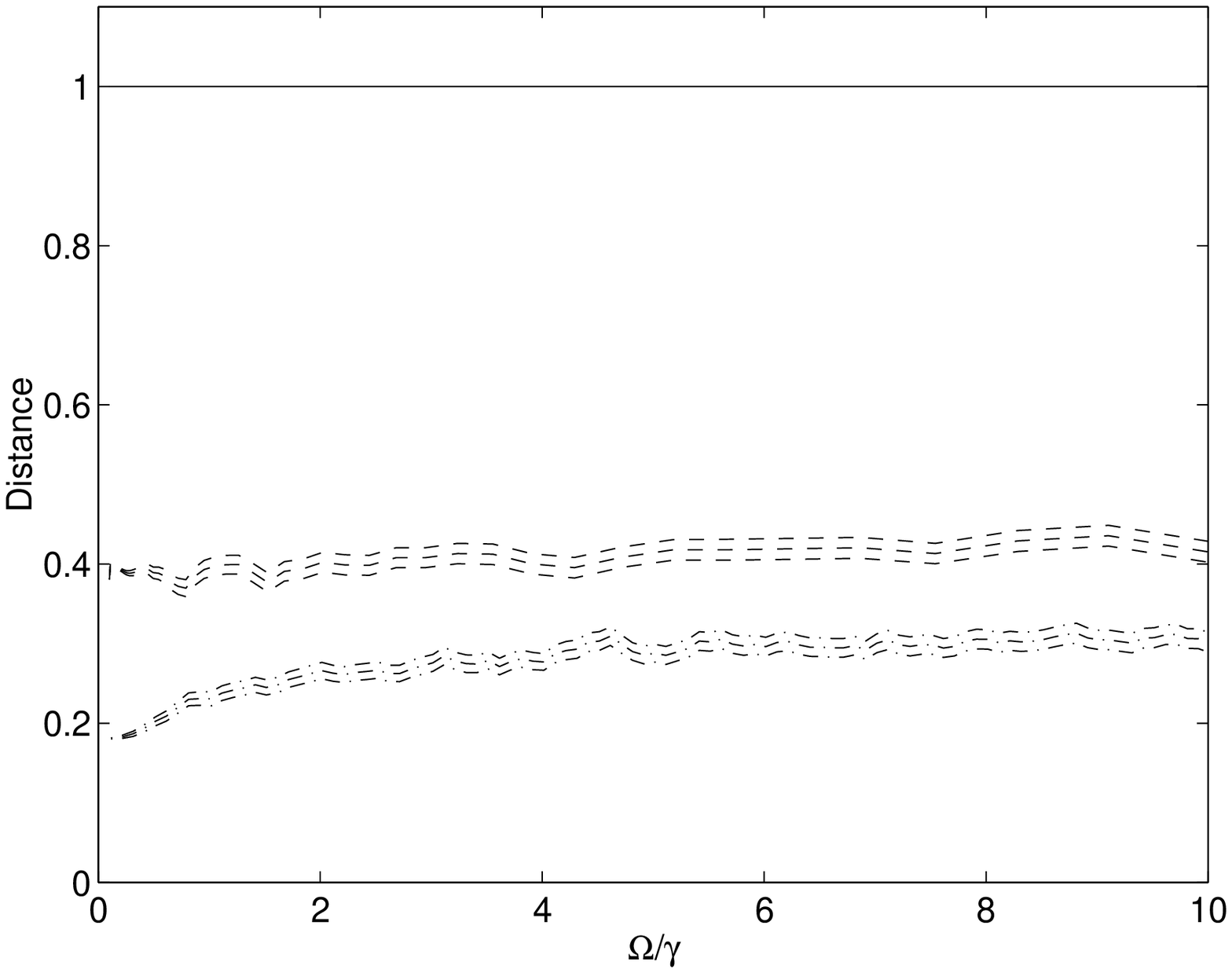}
	\caption{\narrowtext Plot of the distance of various ensembles from 
	the maximally robust ensemble as a function of $\Omega/\gamma$. The 
	solid line is for direct detection, which is the same as for 
	continuous Markovian unraveling (CMU) with $\upsilon=-1$. The dashed line is for 
	a CMU with $\upsilon=0$. The dash-dot  line is for $\upsilon=1$ (the maximally robust 
	CMU). In the latter two cases, the three lines indicate the mean, 
	and the mean plus or minus one standard deviation (the statistical error 
	in the ensemble average). }
	\protect\label{fig:dist}
\end{figure}

Thus to find how close the MRCMU ensemble is to the MRU ensemble we 
simply need to evaluate ${\rm E}[|u|]$, 
the ensemble average of $|\an{\sigma_{x}}|$ 
for the former. The result is plotted in Fig.~\ref{fig:dist}. The 
distance is always less than about $0.3$, the value to which it 
appears to asymptote for large $\Omega$. Since this is moderately 
small compared to one, we can say that the two ensembles are 
moderately close. This is in stark contrast to either the direct 
detection ensemble or the $\upsilon=-1$ CMU ensemble. 
For these two ensembles, $u=0$ for all members, so the distance 
to the MRU ensemble is the maximum value of unity. 

Also plotted in 
Fig.~\ref{fig:dist} is the distance from the CMU with $\upsilon=0$ to the 
MRU ensemble. This CMU has a number of special properties and is 
sometimes known as quantum state diffusion 
\cite{GisPer92a,GisPer92b}. We see that it also 
gives an ensemble whose distance to the MRU ensemble
is less than unity. However, the distance is considerably greater 
than that of the MRCMU, asymptoting to a 
value greater than $0.4$.

\section{Discussion}

The resonance fluorescence master equation generates surprisingly  
rich dynamics for a two-level atom. Here we have investigated those 
dynamics using the technique of finding the maximally robust unraveling. 
That is, finding the scheme for monitoring the fluorescent radiation 
that collapses the atom into pure states that are, on average, the 
most robust. By robust we mean that they survive best  
under the master equation evolution so that, once the monitoring 
ceases, the probability for the atom to be found at some later time to 
still be in  
the state into which it was collapsed by the monitoring, is maximized.

The property of producing robust states may give the maximally robust 
unraveling potential applications, particularly in quantum 
information technology \cite{Bar96,KniLafZur98} where minimizing the 
effect of environmental decoherence is essential. Quite separately 
from any application, the maximally robust unraveling is useful for 
characterizing the dynamics of open quantum systems, such as the 
resonantly driven atom of this study. It has been suggested before 
\cite{WisVac98,WisVac00} 
that the ensemble arising from the MRU is the most natural 
representation of the system's stationary state matrix in terms of 
pure states (state vectors).

In this work we have found that the MRU for the RF 
master equation is an adaptive interferometric monitoring scheme 
proposed in Ref.~\cite{WisToo99}. The atom's radiation is, prior to 
detection by a photodetector, interfered 
at a beam splitter with a reflected local oscillator. The measurement is 
adaptive because the local oscillator amplitude (which is comparable 
in magnitude to the field radiated by the atom) has its phase changed 
by $\pi$ every time a detection occurs. This detection 
scheme has the remarkable property that, in steady state, the atom 
simply jumps between two fixed pure states. In the large driving 
limit, these two pure states are close to eigenstates of the driving 
Hamiltonian $\Omega \sigma_{x}/2$.

The adaptive interferometric monitoring scheme was designed in 
Ref.~\cite{WisToo99} specifically to have this property of producing a 
stationary ensemble containing just two members. In that reference it 
was found that other detection schemes, such as spectral detection 
(resolving the three Mollow peaks), and another adaptive scheme, give 
rise to similar behaviour. That is, the atom jumped between states 
that were close to $\sigma_{x}$ eigenstates in the large driving limit. 
In Ref.~\cite{WisToo99} it was speculated that this behaviour was  
what the atom ``wanted to do''. Here we have confirmed that the 
resulting two-member ensemble  is indeed the most robust and hence 
arguably the most natural. In the context of the study of decoherence 
and the classical limit, it appears that jumping 
between two fixed states is the most classical behaviour for a 
strongly-driven two level atom.

Another issue we have investigated in this paper is how close to the 
MRU one can approach if one restricts the unravelings to continuous 
Markovian ones. In the context of the fluorescent atom, this means 
unravelings realizable from homodyne measurements. They give rise to 
evolution on the Bloch sphere that is continuous (but not 
differentiable) and Markovian. This is an 
interesting question because the set of continuous Markovian unravelings 
is easily parameterized by real numbers, unlike the set of all 
possible unravelings,
which is too large to be finitely parameterized in this way. For this 
reason, previous work in MRU \cite{WisVac98,WisVac00} has concentrated on 
finding the most robust CMU.

In this work we have found that the MRCMU has a robustness, as 
measured by the survival time, which falls as 
$\Omega^{-1}$ as the driving $\Omega$ increases. This is similar to 
the result for direct detection, and contrary to that of the MRU for 
which the survival time is constant at $2\ln 2 \gamma^{-1}$. 
 However, as we have shown graphically, the distribution of states on 
 the Bloch sphere for the MRCMU ensemble is qualitatively much closer to 
 that of the MRU than to that of direct detection. 
   Furthermore, we have introduced a quantitative measure 
for the closeness of two ensembles of pure states, and applied this to 
the various unravelings. We find that the MRCMU ensemble is reasonably 
close to the MRU ensemble, while the direct detection ensemble 
is as distant as is possible from 
the most robust ensemble.

This result has wider implications in the program of decoherence, 
robustness, and the classical limit. A two-level atom is an extremely 
non-classical system. In the limit of strong driving the stationary 
state matrix is almost fully mixed, and the existence of two discrete 
levels manifests strongly in the dynamics: the maximally robust 
unraveling has jumps between two almost orthogonal states. By 
contrast, under a continuous Markovian 
unraveling the atomic state does not jump, but rather diffuses around 
the Bloch sphere. 

We have shown that, despite the atom's nonclassicality, 
the ensemble generated by 
a CMU can be reasonably close to that generated by 
the MRU. This suggests that restricting an investigation to 
continuous Markovian unravelings is not a serious restriction 
(provided that one is not interested in the numerical value of the
 maximum survival time). The 
basic nature of the maximally robust unraveling should be evident 
from that of the maximally robust CMU. For the two-level atom it is 
fortunate that the absolute maximally robust unraveling can be found 
analytically. For more general systems a numerical search would be 
necessary, and, to be practical, would have to be confined to 
finitely parameterizable unravelings such as the continuous Markovian 
unravelings. Thus our results  lends credence to the 
whole program of finding the (approximately) 
maximally robust unraveling for open 
quantum systems.

\appendix
\section*{Derivation of Eq.~(4.7)}

Let $n(t)$ be the event that there were no detections from time 
$t_{0}$ to 
the present time $t_{0}+t$. 
Let $d(t)$ be the event that there was a detection at time $t$ before 
the present.
Then 
\beq
P_{0}(t) = P[n(t)|d(t)],
\eeq
where $P_{0}(t)$ is as in \erf{defP0} and where 
$P[A|B]$ means the probability of $A$ given $B$.

Now what we want is $\wp(t)$, the probability that the last detection 
was at a time $t$ before the present, at steady state. 
That is,  the probability that 
there was a detection at time $t$ in the past, {\em and} 
that there were no detections 
from then until now. In other words,
\beq
\wp(t) = P[n(t) \wedge d(t)],
\eeq
where $P[A\wedge B]$ means the joint probability of $A$ and $B$.

Now from the definition of conditional probability,
$P[A\wedge B] = P[A|B]P[B].$
Therefore
\beq \wp(t) = P_{0}(t)P[d(t)].
\eeq
But at steady state (that is, after initial transients have decayed), 
the probability that there was a detection at time $t$ in the past, given no 
other information, does not depend on $t$. That is, $P[d(t)]$ is a 
constant, so we simply have
\beq \wp(t) =  {\cal N}^{-1} P_{0}(t),
\eeq
for some constant ${\cal N}$.

To find the constant of proportionality we just note that, since the  
last detection must have been at some time $t$ in the past, 
$\int_{0}^{\infty} \wp(t) dt = 1.$
Here we are actually treating $\wp(t)dt$  as the probability that the 
last detection was in the interval $[t,t+dt)$. From this condition it 
is easy to see that 
\beq{\cal N} = \int_{0}^{\infty} P_{0}(t) dt,\eeq
which gives \erf{defwp}.

\acknowledgments
This work was partly supported by the Australian Research Council. HMW 
would like to acknowledge ongoing discussions with John Vaccaro.

\end{multicols}

\begin{multicols}{2}


\begin{figure}
{\bf Figure 1 is attached at the end.}\\
\caption{\narrowtext Plots (a), (c), and (e) 
show the distribution of pure states in the 
stationary ensemble on the 
Bloch sphere under various detection schemes. Plots (b), (d), and (f)
show the decay towards the stationary state matrix of a typical member 
of each ensemble. Plots (a) and (b) are for direct detection, (c) and 
(d) for the maximally robust unraveling, and (e) and (f) for the 
maximally robust continuous Markovian unraveling. For all plots 
$\Omega/\gamma=10$. The ensemble in (c) consists of just two members, 
and the dots indicating their positions have been enlarged to make 
them more easily visible.}
	\protect\label{fig:bs}
\end{figure}

\end{multicols}


\begin{references}


\bibitem{Gea90}
J. Gea-Banacloche,
in: {\em New Frontiers in Quantum Electrodynamics and
Quantum Optics},
A.O. Barut, ed., Plenum, New York (1990).

\bibitem{Zur93}
W.H. Zurek,
Prog. Theor. Phys. {\bf 89}, 281 (1993).

\bibitem{ZurHabPaz93}
W.H. Zurek, S. Habib, and J.P. Paz,
Phys. Rev. Lett {\bf 70},1187 (1993).

\bibitem{Gal95}
M.R. Gallis,
Phys. Rev. A {\bf 53}, 655 (1996).

\bibitem{BarBurVac96}
S.M. Barnett, K. Burnett and J.A. Vaccaro,
J. Res. Natl. Inst. Stand. Technol. {\bf 101},593 (1996).

\bibitem{ParScu98}
Gh.-S. Paraoanu and H. Scutaru
Phys. Lett. A {\bf 238}, 219 (1998).

\bibitem{Gea98}
J. Gea-Banacloche,
Found. Phys. {\bf 28}, 531 (1998).

\bibitem{WisVac98}
H.M. Wiseman and J.A. Vaccaro,
Phys. Lett. A {\bf 250}, 241 (1998).

\bibitem{WisVac00}
H.M. Wiseman and J.A. Vaccaro, 
quant-ph/9906125; submitted to Phys. Rev. A.

\bibitem{note1}
The continuity and Markovicity of the stochastic evolution of the state 
vector under an unraveling is quite distinct (and not implied by) the 
continuity and Markovicity of the ensemble average evolution described by 
a Lindblad master equation.

\bibitem{Gar91}
C.W. Gardiner,
{\em Quantum Noise}
(Springer, Berlin, 1991).

\bibitem{Lin76}
G. Lindblad,
Commun. math. Phys. {\bf 48}, 199 (1976).

\bibitem{Car93b}
H.J. Carmichael,
{\em An Open Systems Approach to Quantum Optics}
(Springer-Verlag, Berlin, 1993).

\bibitem{DalCasMol92}
J. Dalibard, Y. Castin and K. M\o lmer,
Phys. Rev. Lett. {\bf 68}, 580 (1992).

\bibitem{GarParZol92}
C.W. Gardiner, A.S. Parkins, and P. Zoller,
Phys. Rev. A {\bf 46}, 4363 (1992).

\bibitem{WisMil93c} 
H.M. Wiseman and G.J. Milburn,
Phys. Rev. A {\bf 47}, 1652  (1993).

\bibitem{Gar85}
C.W. Gardiner,
{\em Handbook of Stochastic Methods}
(Spring\-er, Berlin, 1985).

\bibitem{QSOSQO96}
Quant. Semiclass. Opt. {\bf 8} (1) (1996),
special issue on ``Stochastic quantum optics",
edited by H.J. Carmichael.

\bibitem{CrePC}
J. Cresser,
private communication.

\bibitem{WisToo99}
H.M. Wiseman and G.E. Toombes,
 Phys. Rev. A {\bf 60}, 2474 (1999).

\bibitem{GisPer92a}
N. Gisin and I. Percival,
Phys. Lett. A {\bf 167}, 315 (1992).

\bibitem{GisPer92b}
N. Gisin and I. Percival,
J. Phys. A {\bf 25}, 5677 (1992).

\bibitem{Bar96}
A. Barenco, 
Contemp. Phys. {\bf 38}, 357 (1996).

\bibitem{KniLafZur98}
E. Knill, R. Laflamme, and W.H. Zurek, 
Nature {\bf 279}, 342 (1998).

\end{references}
\end{document}